\documentclass[aps,prl, twocolumn, superscriptaddress, showpacs,
  showkeys, letter]{revtex4}
\usepackage{graphicx}
\usepackage{textcomp}
\usepackage{color}
\usepackage{dcolumn}
\usepackage{multirow}
\usepackage{ulem}
\usepackage{textcomp}
\begin{document}

\title{Quantum state tomography of an itinerant squeezed
    microwave field}

\author{F.\ Mallet}
\affiliation{JILA, National Institute of Standards and Technology
and the University of Colorado, Boulder, CO 80309, USA}
\author{M.\ A.\ Castellanos-Beltran}
\affiliation{JILA, National Institute of Standards and Technology
and the University of Colorado, Boulder, CO 80309, USA}
\affiliation{Department of Physics, University of Colorado,
Boulder, CO 80309, USA}
\author{H.\ S.\ Ku}
\affiliation{JILA, National Institute of Standards and Technology
and the University of Colorado, Boulder, CO 80309, USA}
\affiliation{Department of Physics, University of Colorado, Boulder,
CO 80309, USA}
\author{S. Glancy}
\affiliation{National Institute of Standards and Technology,
Boulder, Colorado 80305, USA}
\author{E. Knill}
\affiliation{National Institute of Standards and Technology,
Boulder, Colorado 80305, USA}
\author{K. D. Irwin}
\affiliation{National Institute of Standards and Technology,
Boulder, Colorado 80305, USA}
\author{G. C. Hilton}
\affiliation{National Institute of Standards and Technology,
Boulder, Colorado 80305, USA}
\author{L. R. Vale}
\affiliation{National Institute of Standards and Technology,
Boulder, Colorado 80305, USA}
\author{K. W. Lehnert} \email{konrad.lehnert@jila.colorado.edu}
\affiliation{JILA, National
Institute of Standards and Technology and the University of
Colorado, Boulder, CO 80309, USA} \affiliation{Department of
Physics, University of Colorado, Boulder, CO 80309, USA}

\date{\today}

\begin{abstract}
We perform state tomography of an itinerant squeezed state of the
microwave field prepared by a Josephson parametric amplifier (JPA). We
use a second JPA as a pre-amplifier to improve the quantum efficiency
of the field quadrature measurement (QM) from 2 \% to
$36\pm4~\%$. Without correcting for the detection inefficiency we
observe a minimum quadrature variance which is $68^{+9}_{-7}~\%$ of
the variance of the vacuum. We reconstruct the state's density matrix by a maximum
likelihood method and infer that the squeezed state has a minimum variance less than 40
\% of the vacuum, with uncertainty mostly caused by calibration
systematics.
\end{abstract}


\maketitle

Fundamental quantum optics experiments at microwave frequencies have
been recently performed with superconducting qubits or Rydberg atoms
inside high-quality microwave cavities. Examples include the
reconstruction of the Wigner functions of Fock states from one
\cite{houck2007} to a few photons and coherent superpositions of few
photons \cite{Deleglise2008,hofheinz2008,hofheinz2009}.  States such
as these, which are manifestly nonclassical light states, are crucial
for quantum information processing, because they can be used to
generate entanglement. However, in the cited experiments, these states
are confined in cavities.  Therefore distributing entanglement to
separate parties, as required in quantum communication protocols,
remains challenging for microwave implementations. In contrast to the
discrete Fock state approach, continuous variables quantum information
(CVQI) strategy uses another type of nonclassical states, the squeezed
states, which are readily created in itinerant modes. These states
exhibit reduced noise, below the vacuum fluctuations, in one of their
quadrature components and amplified noise in the other one. They are
also easily generated at optical frequencies in the itinerant output
modes of parametric amplifiers made of optically nonlinear
crystals. At optical frequencies, CVQI has progressed rapidly from the
initial creation of squeezed states \cite{Slusher1985} and tomographic
reconstruction \cite{Smithey1993,Schiller1996,Breitenbach1997} of
those states to teleportation \cite{furusawa1998,yonezawa2007} and
quantum error correction \cite{aoki2009,lassen2010}.

At microwave frequencies, the field is less advanced. The generation
of microwave squeezed states using the nonlinear electrical response
of superconducting Josephson junctions has been reported
\cite{Yurke1988}, with inferred squeezing down to $10~\%$ of vacuum
variance \cite{castellanos-beltran2008}.  Such states can be powerful
tools for quantum information processing and communication because
microwaves and superconducting qubits can mimic useful light--atom
interactions, as demonstrated in \cite{Wallraff2004}.  Furthermore,
these devices are made of compact and integrable electrical circuits,
with much promise for building complex quantum information
processors. The lack of an efficient quadrature measurement (QM) for
itinerant modes has slowed the advancement of CVQI. However, as
demonstrated recently in \cite{teufel2009}, it is possible with a JPA
to realize an efficient single QM.

In this Letter, we report the tomography of an itinerant squeezed
microwave field. We demonstrate that our JPA based measurement scheme
has a quantum efficiency $20$ times greater than a QM employing
state-of-art semiconductor amplifiers. We infer the quantum state
prepared by maximum likelihood tomography, correcting for inefficiency
in our QM.  We discuss the achieved degree of squeezing, from the
perspective of generating entanglement on chip.

\begin{figure}
 \includegraphics[width=75mm]{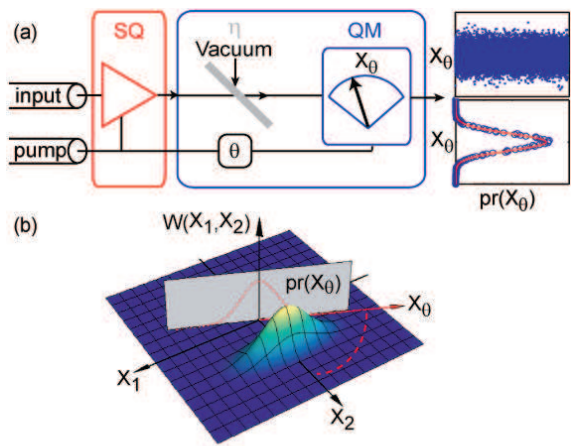}
\caption{\label{fig_principle} Principle of the experiment. ~(a-left):~The
squeezer (SQ, in red) prepares a squeezed state whose quadrature
distributions are measured for different phases $\theta$ with an
efficiency $\eta$. ~(a-right):~Simulated measurement results for
20,000 realizations of creating the squeezed state and measuring it
at a single $\theta$. The top graph shows the measured quadrature
value versus realization number. The bottom plot is a histogram
(blue circles) and Gaussian probability distribution
$\textrm{pr}(X_{\theta})$ (red curve) of
this random process.  ~(b):~Graphical
Interpretation: the probability distribution
$\textrm{pr}(X_{\theta})$ is simply the projection of the Wigner
function%: $\textrm{pr}(X_{\theta})=\int dX_{\theta+\pi/2}
%W(X_\theta,X_{\theta+\pi/2})$
.}
\end{figure}

Homodyne tomography is a standard experimental tool to infer the
quantum state of a single mode of light.  It was proposed in
\cite{vogel1989} and pioneered on a squeezed optical field in
\cite{Smithey1993}.  Its principle is depicted in the
Fig.~\ref{fig_principle}.  A homodyne detection apparatus measures the
value of the quadrature $X_{\theta}$, where $\theta$ is set by
adjusting the phase of the local oscillator. The probability density
function $\textrm{pr}(X_{\theta})$ for
measuring a particular value of $X_\theta$ is the marginal density function of the Wigner function, i.e. $\textrm{pr}(X_{\theta})=\int dX_{\theta+\pi/2}
W(X_\theta,X_{\theta+\pi/2})$, as shown in
Fig.~\ref{fig_principle}~(b).  Thus by performing measurements of
$X_\theta$ on many identical copies of the %\sout{system}
 state and varying
$\theta$, the ``hidden'' quantum object can be seen from different
angles and its state inferred. Losses and other Gaussian noise sources
in the homodyne detector can be modeled with the insertion of a
fictitious beam splitter of transmissivity $\eta$, as shown in
Fig.~\ref{fig_principle}~(a).  In such a case, the measured $\textrm{pr}(X_{\theta})$ are no
longer the projections of the desired Wigner function $W$, but of a smoother
%\sout{quasi-probability}
distribution
%\sout{created by}
which is the convolution of
%\sout{the desired Wigner function}
$W$ with a Gaussian Wigner function \cite{leonhardt1993}. However methods like maximum
likelihood quantum state tomography can be used to deconvolve the effect
of inefficiency \cite{lvovsky2005}.

At optical frequencies, $\eta\ge90~\%$ is routinely obtained using a
pair of balanced photodiodes \cite{lvovsky2005}. Such detectors are
not available for microwaves and until recently the best setup was a
chain of phase-insensitive amplifiers followed by a mixer, or two such chains in parallel \cite{menzel2010,mariantoni2010,bozyigit2010}. In
such a case, noise $A_n$ greater than $1/2$ (the vacuum variance) must be added to the QM \cite{caves1982}.
This noise can be modeled as an effective efficiency by the relation
$\eta=1/(1+2A_n)$ \cite{leonhardt1994}, so the QM efficiency using
phase-insensitive amplifiers is limited to $50~\%$.  State of the art microwave amplifiers,
high-electron-mobility transistors (HEMTs), have $A_{n}\approx10-20$.
In practice, the unavoidable losses present in a microwave experiment
typically result in $\eta \approx 2$\%. However, as demonstrated in
\cite{teufel2009}, inserting a JPA used as a single quadrature
preamplifier before the HEMT increases the experimentally achieved
$\eta$ by a factor of approximately 20.

\begin{figure}%[!ht]
 \includegraphics[width=75mm]{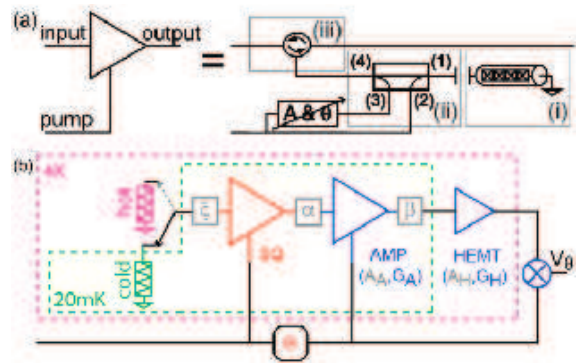}
\caption{\label{fig_schematics} ~(a):~To implement a SQ or AMP at
  microwaves, three microwaves components are required: (i) a JPA,
  (ii) a directional coupler and (iii) a circulator. Taking port
  (1) of the directional coupler as reference, (2) is the weakly
  coupled port, (3) the isolated port and (4) the direct port. Port (2) is
  used to pump the JPA. Port (3) is used to apply a cancelation tone (adjusted with a room temperature
  attenuator and phase shifter) that nulls the pump and
  displaces the output of the JPA back to the origin of the phase space.~(b):~Schematic of the experiment. In
  this figure, all the microwave components and cables are considered
  lossless; their imperfections are absorbed into the
  experimentally determined total transmissivities $\xi$, $\alpha$ and $\beta$.}
\end{figure}

To perform a high-quality reconstruction of the Wigner function of a
squeezed microwave state we operate two JPAs in series, as shown in
Fig.~\ref{fig_schematics}~(b). The first JPA, referred to as the
squeezer (SQ), prepares the squeezed state. The second JPA, referred
to as the pre-amplifier (AMP), amplifies the quadrature of the
squeezed state determined by the phase difference $\theta$ between the
AMP and the SQ pump tones. We vary $\theta$ by applying to the two cavities pump tones slightly detuned
from one another. The SQ stage is pumped at $7.45$~GHz,  while the AMP stage
is pumped at $100$~kHz higher frequency; therefore, sweeping $\theta$ through
$2\pi$ every $10$~\textmu s.

Our implementation of an SQ or an AMP at microwaves, as shown in
Fig.~\ref{fig_schematics}~(a), requires three elements: (i) a JPA used
in reflection, (ii) a directional coupler and (iii) a circulator.  As
described in~\cite{castellanos-beltran2008}, the JPAs are nonlinear resonant cavities built from coplanar
waveguides whose central conductor has been replaced by a series of
many Josephson junctions. The Josephson junctions' nonlinearity
causes the cavity's phase velocity to be intensity
dependent. Therefore when the cavity is pumped it becomes a
phase sensitive amplifier for input modes whose frequencies lie
within the bandwidth of the JPA centered on the pump frequency.
Such microwave modes incident on the JPA are reflected and exit the
cavity with one quadrature amplified and the other squeezed, depending on
their phase relative to the pump's phase.
A directional coupler is used to add
the pump tone to the incident signal and remove the pump tone from the reflected signal.
%
%A directional coupler
%allows the pump to enter through the weakly coupled port while the
%reflected signal simply flows through the direct port. The
%reflected signal consists of amplified and squeezed modes displaced
%from the origin of phase space by the reflected pump tone.  That
%displacement is approximately removed by applying the appropriate
%tone to the isolated port of the directional coupler, so that the
%reflected pump tone is canceled by interference.
Finally the incident
and reflected modes are separated into different cables using a
circulator.

Following Fig.~\ref{fig_schematics}~(b), in the limit of large HEMT
power gain $G_{H}$, our quantum efficiency can be cast as
\begin{equation}
\label{efficiency}
%\eta=\frac{\alpha}{1+2(A_{A}+(1-\alpha)/2+(A_{H}+(1-\beta)/2)/G_{A}\beta)},
\eta=\frac{\alpha}{2+2A_A - \alpha + [2A_H - (1- \beta)]/G_{A}\beta},
\end{equation}
where $A_A$ ($A_H$) is the AMP (HEMT) added noise, $\alpha$ ($\beta$)
is the fraction of power transmitted by the microwave circuitry
between the SQ and the AMP (the AMP and the HEMT), and $G_A$ is the
power gain of the AMP stage.  A detailed description of how we
calibrate each of these parameters is in the supplementary
information.
%%
%%
%%Briefly, we infer the added noise and loss of the elements by injecting one
%%of two sources of known microwave noise--selected by connecting a microwave
%%switch to microwave termination held either a base temperature (cold load) or
%%4.1~K (hot load)-- and observing the change in the noise of the QM.
%%%%
%%
%The general strategy is to put thermal noise at
%different temperatures into the different segments of the circuit in
%Fig.~\ref{fig_schematics}, operate each JPA either as an amplifier
%(ON) or as a noiseless element with unit gain (OFF), and observe
%the variation in the noise at the output of the measurement
%chain.  A microwave switch connects the input of the SQ to either a
%``hot load'' (50 $\Omega$ microwave termination at $4.1$~K) or a
%``cold load'' (at $20$~mK), preparing the modes in the cable in the
Briefly, we inject different
amounts of thermal noise into the amplifier chain while operating each JPA either as an amplifier
(ON) or as a noiseless element with unit gain (OFF). We then infer the added noise and loss of the elements by
observing the variation in the noise at the output of the measurement
chain. The thermal noise is varied by connecting the input of the SQ through a switch to either a
``hot load'' (50~$\Omega$ microwave termination at $4.1$~K) or a
``cold load'' (at $20$~mK).
Although the tomography  is only performed with the ``cold load'', both are required for
calibration.
We obtain $A_A=0.25\pm0.06$, $A_H=17.3\pm0.1$,
$\alpha=68\pm 2~\%$ and $\beta=74\pm5~\%$.  However, as the
switch is operated at the 4.1 K stage and is slightly lossy, the state presented at
the input of the SQ with the ``cold load'' is not pure quantum vacuum,
but a low occupancy thermal state with average photon number
$\overline{n}\simeq0.15 \pm 0.15$. One quadrature of the resulting squeezed state is then amplified at the
AMP stage with sufficient gain $G_A=180$ such that the noise in the
amplified quadrature exceeds $A_H$ for any $\theta$. From
Eq.~(\ref{efficiency}), we obtained an overall quantum efficiency of
$36\pm 4 ~\%$, which can be compared to $\eta\approx 2 ~\%$ without
the AMP stage.

In this experiment our uncertainty in $\eta$ and $\overline{n}$ create
a systematic source of error.  We thus perform our data analysis under
three assumptions (1) high efficiency ($\eta=0.40$) and high mean
photon number ($\overline{n}=0.30$), (2) best estimate for both
efficiency ($\eta=0.36$) and mean photon number ($\overline{n}=0.15$),
and (3) low efficiency ($\eta=0.33$) and low mean photon number
($\overline{n}=0$).  These three cases give us ``pessimistic'',
``best-guess'', and ``optimistic'' analyses, in terms of the
purity of the squeezed state estimated by the tomography. Using a
lower estimate for $\eta$ and $\overline{n}$ as inputs to the
tomography algorithm causes it to return a more pure, more squeezed, and therefore a more ``optimistic''
estimate of the squeezed state.  Associated with each of these
three cases, we also have statistical uncertainty, so the given error
bounds cover an interval that includes both uncertainties around the
``best-guess'' estimate. They are reported
in the form $X_{-L}^{+U}$, where $X$ is the statistical mean using the ``best-guess''
calibration and $L$ and $U$ are respectively the lower and upper bounds of
the one standard deviation uncertainty in the ``pessimistic'' and
``optimistic'' cases.

\begin{figure}
\includegraphics[width=82mm]{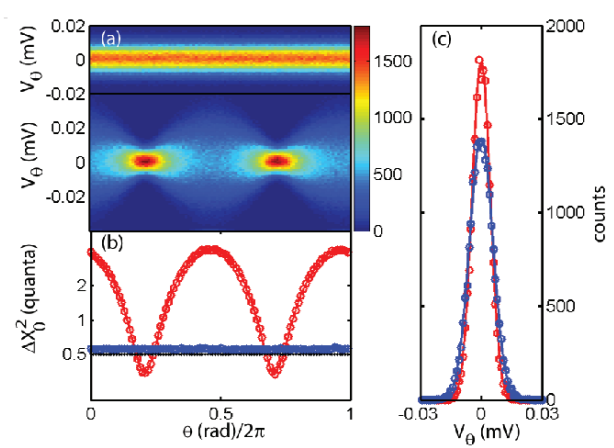}
\caption{\label{NoiseVsPhase}~(a):~Density plot of number of occurrences in a  $1~$\textmu V bin size of the amplified quadrature voltage $V_\theta$  versus $\theta/2\pi$, with the SQ pump OFF (top) and ON (bottom).~(b):~In particular, histograms of $V_\theta$ at the
maximum of squeezing: data ($\circ$) and Gaussian fit (continuous
lines) for the SQ pump OFF (blue) and ON (red).~(c):~Noise variance
$\Delta X^2_{\theta}$ in quanta units on a log scale versus
$\theta/2\pi$ for the SQ pump ON (red) and OFF (blue).  The (black)
 line indicates our estimate of the vacuum noise level under the ``best-guess'' calibration.}
\end{figure}

We must also calibrate the QM to convert the measured voltage noise into units of noise quanta. In optical
homodyne tomography, this is usually done by inserting the vacuum and
observing the quadrature noise. Analogously, we insert the weak
thermal state with mean photons $\overline{n}$ (by
simply turning the SQ stage OFF) and measure voltages proportional to
quadrature values at many $\theta$, as shown (in blue) in
Fig.~\ref{NoiseVsPhase}. As expected this voltage noise is $\theta$
independent, with a variance $\Delta V_{\mathrm{ SQ,OFF}}^2=3.2\times
10^{-5}~\textrm{mV}^2$. Under the convention that vacuum has variance
$1/2$ in unitless quadrature space (or in units of ``quanta''), we
calibrate this voltage variance to $\Delta X_{\mathrm{SQ,OFF}}^2 =
(1-\eta)/2+\eta(1/2+\overline{n}) = 0.55_{-0.05}^{+0.07}$ quanta.
Therefore the desired conversion factor  $\Delta X_{\mathrm{SQ,OFF}}^2/\Delta
V_{\mathrm{SQ,OFF}}^2=1.71_{-0.17}^{+0.20} \times 10^4~\mathrm{quanta/mV}^{-2}$
is used to rescale the variances in Fig.~\ref{NoiseVsPhase}~(c).

In Fig.~\ref{NoiseVsPhase}~(a), we show QM data of the squeezed state.
With SQ ON (red) we observe
the characteristic phase dependent noise for a squeezed state.  At the
phase for which the variance is minimum, we show the histogram of
quadrature measurements in Fig.~\ref{NoiseVsPhase}~(b). The SQ OFF
histogram is clearly wider than the SQ ON histogram, demonstrating our
ability to observe squeezing directly at the output of our measurement
chain.  In Fig.~\ref{NoiseVsPhase}~(c) we plot the variance of the QM
with SQ ON and OFF as a function of $\theta$, expressed in units of
quanta, clearly showing squeezing below the vacuum level.  Without
correcting for $\eta$, we observe a minimum quadrature variance which
is $\Delta X^2_{\mathrm{SQ,MIN}}=68_{-7}^{+9}~\%$ of the vacuum variance.
%Using the equation $V_{\mathrm{obs}}=\eta V_{\mathrm{sq}} +
%\left(1-\eta\right)/2$ we can obtain a linear unbiased estimate of
%$V_{\mathrm{sq}}$, the squeezed state's variance corrected for loss
%$1-\eta$.
%We find the minimum $V_{\mathrm{sq}}=10^{+30}_{-10}~\%$ of
%the vacuum variance.

\begin{figure}
\includegraphics[width=80mm]{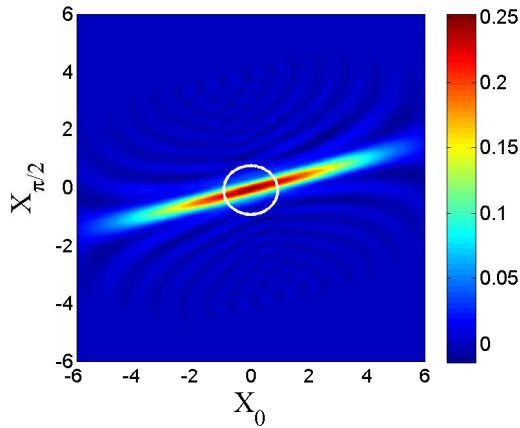}
\caption{\label{Wigner}Mean of 35 reconstructions of the Wigner function of the state exiting the SQ,
  inferred by maximum likelihood under the ``best-guess''
 assumption, in quanta units. The faint pattern of ripples extending from
  the origin is caused by truncation at 30
  photons of the density matrix used to represent the state. The white circle at the origin
  shows the full-width at half-maximum of the vacuum state.}
\end{figure}

To infer the quantum state created by the squeezer, correcting for
loss during the QM, we used maximum likelihood quantum state
tomography \cite{paris2004}.  For each of the three calibration cases,
we performed 35 reconstructions using independent subsets each
containing 10,000 QMs of the total measured data.  We estimated
statistical uncertainty from the spread of properties (such as
fidelity or minimum variance) of the set of 35 reconstructions.  The
statistical uncertainty was significantly lower than the systematic
uncertainty.
%Our estimates of the squeezed state's properties
%are reported in the form $X_{-L}^{+U}$, where $X$ is the mean of the
%35 maximum likelihood estimates based on the ``best-guess''
%calibration parameters.  $L$ and $U$ are the lower and upper bounds of
%the one standard deviation uncertainty interval associated with our
%maximum likelihood estimate based on the ``pessimistic'' and
%``optimistic'' cases.
In Fig.~\ref{Wigner} we show the Wigner
function of the ``best-guess'' reconstructed state $\rho$. The \textit{pure} squeezed
vacuum state $|\psi\rangle$ that has the highest fidelity with $\rho$ has minimum
quadrature variance $6.0_{-1.1}^{+1.4}~\%$ of
the vacuum variance, and that maximum fidelity is
$F=\langle\psi|\rho|\psi\rangle = 0.81_{-0.17}^{+0.16}$. As explained in the supplementary information, the minimum variance of $\rho$ is biased by an amount comparable to our systematic uncertainty, so we infer the minimum variance $\Delta x^2_{\mathrm{SQ,MIN}}$ directly from the observed minimum variance as $\Delta x^2_{\mathrm{SQ,MIN}}=(1/\eta)(\Delta X^2_{\mathrm{SQ,MIN}}-(1-\eta)/2)$.
We find $\Delta x^2_{\mathrm{SQ,MIN}}=12^{+30}_{-12}~\%$ of
the vacuum variance.
%
%However, the
%most likely inferred state's squeezed quadrature is actually
%$48^{+45}_{-19}~\%$ of the vacuum's variance, and the anti-squeezed
%quadrature's variance is $20.2^{+5.4}_{-1.0}$ times the vacuum
%variance. The discrepancies among the minimum variances of $\rho$
%and $\left|\psi\right\rangle$ and $V_{\mathrm{sq}}$ are largely
%consistent with those found in Gaussian simulations, as discussed in
%the supplementary material.
%
For comparison, the most highly squeezed optical state
ever made has a variance of only $7~\%$ of the vacuum variance
\cite{mehmet2010}.

Producing squeezed states of itinerant modes allows the
generation of distributable entanglement by sending two copies of a squeezed vacuum
state through the two input ports of a balanced beam splitter.  The coherent information \cite{schumacher1996} is one
useful way to characterize the entanglement between the two output
modes.  The asymptotic number of maximally entangled qubit pairs
(e-bits) that can be distilled per copy of the noisy entangled
state, by using local operations and one-way classical communication,
is at least as large as the coherent information \cite{devetak2005}.
Given two copies of $\rho$, one
could make two entangled modes with $2.5^{+1.0}_{-0.4}$ e-bits of
coherent information.

In conclusion, we have reconstructed the Wigner function of an
itinerant squeezed microwave field generated at the output of a
Josephson Parametric Amplifier. Using a second JPA as a preamplifier
has increased the quantum efficiency of the microwave homodyne
detection from approximately $2~\%$ to $36~\%$. The level of
squeezing is primarily limited by noise added to the squeezed state
by the JPA.  Improving the performance of the JPAs (as both
squeezers and phase-sensitive amplifiers) will require more
detailed investigation of the source of this noise. We used maximum
likelihood quantum state tomography to deconvolve the QM
inefficiency in order to precisely characterize the state generated.
This is an important step toward generating easily distributable
microwave entanglement on chip.

\textit{Notes}: A different method was recently used to obtain a similar state reconstruction \cite{eichler2010}.

\begin{acknowledgments}
The authors acknowledge support from the DARPA/MTO QuEST program.
\end{acknowledgments}

\bibliography{journal_titles_abbreviated,mybib2}

%\begin{document}
\onecolumngrid

\section{Supplementary Materials for ``Quantum state tomography of an itinerant squeezed
    microwave field''}

\section{Data acquisition and calibration}
Determining the amplifier added noise and loss requires several calibration steps that permit us to isolate the effect of a specific loss or added-noise contribution to the overall efficiency of the homodyne measurement. The crucial aspect that makes this calibration possible is that the JPA cavities have widely-tunable resonance frequencies, adjusted by imposing a magnetic flux \cite{castellanos-beltran2008,castellanos-beltran2009}. Far from resonance the JPA cavities behave as open circuits. They are simply mirrors that reflect the microwave field without otherwise transforming it; therefore, either the SQ or AMP or both stages can effectively be bypassed.

We begin with both JPA stages bypassed, so that they have $G_{\rm{S}}=G_{\rm{A}}=1$. If the switch were lossless, when it is connected to the cold load, the noise power exiting the HEMT amplifier would be $S=G_{H}(A_H+S_{f})$, where $S_f=(1/2)+n_f=(1/2)+[\exp(\hbar\omega/k_B T_f) -1 ]^{-1}$ and $T_f$ is the refrigerator's temperature. Notice that the result doesn't depend on the transmissivities $\alpha$, $\beta$, or $\xi$ because these are at the same temperature as the cold load, consequently each loss component emits as much power as it absorbs.  However, with the switch connected to the hot load, the expression for the total power at the output becomes $S=G_{H}(A_H + (\xi \alpha \beta)S_h + (1-\xi\alpha\beta) S_f)$, with $S_h=(1/2)+n_h = (1/2)+[\exp(\hbar\omega/k_B T_h) - 1]^{-1}$ and $T_h=4.1$~K. In both cases, we expect and observe that $S$ depends linearly on $S_f$ with an offset. By fitting these linear dependencies we can extract $G_{H}$, $A_H$, and the product $\xi \alpha \beta$.

We cannot assume that the switch is lossless. Because its loss sits at 4.1~K, it will always emit noise power $S_h(1-\lambda)+S_{in}\lambda$, where $S_{in}$ is the incident noise and $\lambda$ is the switch transmissivity. So, even when $n_f\ll1/2$, the state presented at the SQ stage will have average thermal occupancy $\bar{n}=(1-\lambda)\xi n_h$. We write the noise power at the output as function of $S_f$, for switches in both positions as
\begin{equation}\label{eq:AmpAndSqOff1}
  S_{1c} = G_{H}A_H + S_h G_{H}  (1-\lambda)\xi\alpha\beta + S_f[G_{H}\lambda\xi\alpha\beta + G_{H}(1-\xi\alpha\beta)]=b_{1c}+m_{1c} S_f
\end{equation}

\begin{equation}\label{eq:AmpAndSqOff2}
  S_{1h} = G_{H}A_H+S_h G_{H} (\xi\alpha\beta) + S_f[G_{H}(1-\xi\alpha\beta)]=b_{1h}+m_{1h} S_f,
\end{equation}
where the subscript $1c$ ($1h$) corresponds to the switch connected to the cold (hot) load.  Fitting our noise data to the right hand side of  Eq.~\ref{eq:AmpAndSqOff1} and~\ref{eq:AmpAndSqOff2}, we can obtain the four parameters  $b_{1h},b_{1c},m_{1h}$ and $m_{1c}$. However as these parameters are not independent, $S_h=(b_{1h}-b_{1c})/(m_{1c}-m_{1h})$, we cannot extract the switch loss independently. We can nevertheless bound this unknown loss by taking a worst case estimate as the manufacturers minimum specified transmission (at room temperature) $\lambda=0.83$ and assuming it is less lossy at 4.1~K. We moreover confirmed that at room temperature the frequency dependent loss of the switch is within the manufacturer's specification. Then by using $1<\lambda<0.83$, we can bound the desired parameters using Eq.~\ref{eq:AmpAndSqOff1} and~\ref{eq:AmpAndSqOff2}, with the expressions $(\xi\alpha\beta)^{-1}=1+m_{1h}S_h\lambda/(b_{1h}-b_{1c})$, and $G_{H}=m_{1h}/(1-\xi\alpha\beta)$, and $A_H=(b{1c}/G_H) - (1-\lambda)S_h (\xi\alpha\beta)$.

\begin{figure}
\includegraphics[width=7 in]{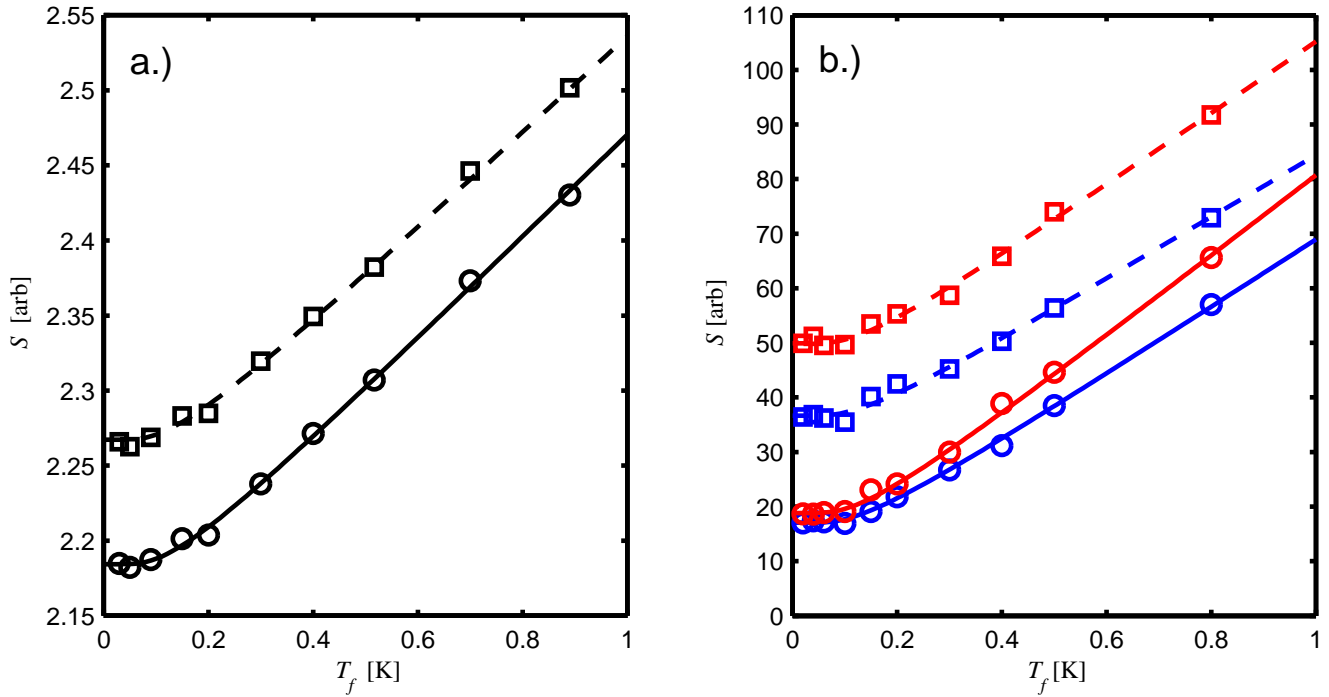}
\caption{\label{fig:SupMatFig1}The noise density $S$ in arbitrary units at the output of the measurement versus refrigerator temperature $T_f$. a.) Data acquired with the AMP and SQ OFF and the switch connected to the cold load (circles) and hot load (squares). The lines are \textit{linear} fits to $S$ versus $S_f$ for the case of the switch connected to the cold load (solid) and hot load (dashed).  b.) Data acquired with the AMP ON and SQ OFF (blue) and AMP OFF and SQ ON (red), with the switch connected to the cold load (circles) with a linear fit (solid) and hot load (squares) with a linear fit (dashed). The arbitrary y-scale is consistent between the six plots. The linear fits do not appear as lines because we plot $S$ versus $T_f$ rather than $S_f$.}
\end{figure}

We then perform the same analysis, finding the linear dependence of the output noise on $S_f$ and on the switch position, with the AMP ON and SQ OFF. From these fits and knowledge of $A_H$ and $G_{H}$ we find $\xi\alpha$, $A_A$, and$G_A\beta$. Finally, we operate the experiment with AMP OFF and SQ ON. A third time we fit the linear dependence of $S$ on $S_f$ with the switch in both positions, determining $\xi$, $\alpha$ and $\beta$ separately (Fig.~\ref{fig:SupMatFig1}b). We evaluate the expressions for $\alpha$, $\beta$, $A_A$, $A_H$, $G_{H}$ and $\bar{n}$ at the bounds on $\lambda$, finding the range of values in the main text. We also find $\xi=-9.9\pm1$~dB, of which 6~dB arises from an attenuator that has been placed at the input of the SQ stage.

To acquire these calibration data sets, we regulate the refrigerator's temperature at 10 values between base temperature ($T < 50$~mK) and 800~mK, which requires about 7 hours to complete. For each temperature point we measure the noise at the output under all six conditions, 2 switch positions, and 3 amplifier configurations (AMP OFF SQ OFF, AMP ON SQ OFF, and AMP OFF SQ ON). We inject a tone detuned from the AMP pump by 20~kHz. By dividing the noise power at the output of the chain by the power in this tone, we become insensitive to any variation in $G_{H}$ over the time needed to acquire the data. At the end of the calibration, we immediately operate the experiment with SQ ON and AMP ON, to acquire the data in the paper. In addition, we use the tone to ensure that we do not saturate the amplifier chain.

Data is acquired by digitizing the output (IF port) of the mixer at rate of $10^{7}$ samples per second. We filter the IF port with a 5~MHz anti-aliasing low-pass filter. The digitized data is digitally filtered with a 3rd-order Butterworth high-pass filter with a 500 kHz corner (3~dB) frequency. The noise density $S$ is the average noise density in the frequency range between 500~kHz and 5~MHz.

\section{Maximum likelihood analysis of the squeezed state}

\begin{table}
\caption{\label{uncertainties} Inferred properties of the squeezed state, upon our three analysis assumptions.}
\begin{ruledtabular}
\begin{tabular}{lccc}
    & Pessimistic & Best guess & Optimistic \\
  \hline
  Fidelity & $0.66\pm 0.02$ & $0.807\pm 0.016$ & $0.960\pm 0.005$ \\
\multicolumn{4}{l}{Min. var. of comparison pure state $\left|\psi\right\rangle$\footnote{Ratio
  of the variance of the squeezed quadrature of the pure squeezed
  vacuum state with highest fidelity to the variance of the vacuum.}:}\\
    & $0.065\pm 0.009$ & $0.060\pm
  0.003$ & $0.0493\pm0.0006$ \\
  $\rho$'s purity & $0.62\pm0.02$ & $0.74\pm0.02$ & $0.96\pm0.01$ \\
  $\rho$'s sq. var.\footnotemark[2] & $0.918\pm 0.002$ & $0.484\pm0.013$ &
  $0.304\pm0.008$ \\
  $\rho$'s anti-sq. var.\footnotemark[2] & $25.54\pm 0.07$ & $20.17\pm 0.06$ &
  $19.18\pm 0.05$ \\
  Coherent info.\footnotemark[3] & $2.19\pm 0.08$ & $2.46 \pm 0.09$ &
  $ 3.42\pm 0.05$ \\
  Linear sq. var.\footnotemark[4] & $0.40 \pm 0.02$ & $0.12 \pm 0.02$
  & $-0.18 \pm 0.02$ \\
\end{tabular}
\end{ruledtabular}
\footnotetext[2]{Ratio of variance of most likely state $\rho$'s squeezed or anti-squeezed
  quadrature to the variance of the vacuum.}
\footnotetext[3]{Coherent information (in e-bits) that could be
  produced with two copies of the squeezed state and a beam splitter.}
\footnotetext[4]{Direct linear inference of the squeezed state's
  minimum variance, relative to vacuum variance.}
\end{table}

Table \ref{uncertainties} shows the statistical errors in our
estimates of inferred parameters characterizing the squeezed state,
for the three analysis cases, based upon our systematic
calibration uncertainties. The first line presents the fidelity
$F=\langle\psi|\rho|\psi\rangle$, where $\rho$ is the maximum likelihood
reconstructed density matrix of the field exiting the SQ and
$|\psi\rangle$ is the pure vacuum squeezed state
  that maximizes the fidelity. The second line gives the ratio of the minimum
  variance of $|\psi\rangle$ to the variance of
  vacuum. The third line gives the purity $\mathrm{Tr}({\rho^2})$ of $\rho.$ The fourth and fifth lines give the ratios of the squeezed
  and anti-squeezed variances of the reconstructed state to the
  variance of vacuum. The sixth line presents the coherent information
  that could be obtained by combining on a beam splitter two copies
  $\rho$. The last line gives our estimate the the experimental
  states' minimum variance based on direct linear inference.

We have stated three variances that characterize the state created
in this experiment: the linear estimate of the experimental state's minimum
variance ($12~\%$), the most likely state $\rho$'s minimum
variance ($48~\%$), and the minimum variance of the pure
squeezed vacuum state $\left|\psi\right\rangle$ that maximizes the fidelity
with $\rho$ ($6.0~\%$).  Here we give more discussion of
these variances.

The quadrature measurements we observe are the linear combination of
the quantum state created by the squeezer and vacuum fluctuations:
\[
X_{\theta}=\sqrt{\eta} x_{\theta}+\sqrt{(1-\eta)}y_{\theta},
\]
where $x_{\theta}$ is the quadrature of the squeezed state, and
$y_\theta$ is the quadrature of the vacuum state.  Solving for
$x_\theta$ gives
\[
x_\theta=\frac{1}{\sqrt{\eta}}\left(X_\theta-\sqrt{\left(1-\eta\right)}y_\theta\right).
\]
Therefore the inferred variance of the squeezed state's quadrature
$\Delta x_\theta^2$ is
\[
\Delta x_\theta^2=\frac{1}{\eta}\left[\Delta X_\theta^2-\left(1-\eta\right)\Delta y_\theta^2)\right].
\]
The vacuum variance $\Delta y_\theta^2=1/2$, and we
can easily calculate an unbiased estimate of
$\Delta X_\theta^2$ for every phase $\theta$.  This
gives us an unbiased estimate of $\Delta x_\theta^2$
that does not depend on the details (for example, Gaussianity) of the
quantum state.  We calculate $\Delta x_\theta^2$ using
20,000 quadrature measurements at each of 100 evenly spaced $\theta$
and calculate the minimum  value $\Delta x_{\mathrm{SQ,MIN}}^2$, in Table
\ref{uncertainties}.  The statistical uncertainties show one standard
deviation in the estimate of $\Delta x_{\mathrm{SQ,MIN}}^2$.  For the ``optimistic
case'' we calculate a negative variance, which is clearly unphysical.
This is a sign of inconsistency in the ``optimistic'' calibration
parameters.  Because the ``optimistic'' estimate for the squeezed
state is computed using the lower bounds on $\eta$ and $\overline{n}$,
this negative variance is evidence that the detector's true $\eta$ and
/ or effective gain ($\Delta X_{\mathrm{SQ,OFF}}^2/\Delta
V_{\mathrm{SQ,OFF}}^2$) must be larger than the lower bounds set by
calibration.

 The minimum variance of $\rho$ is significantly higher than this
  linear estimate.  This is caused by bias in the maximum likelihood
  method.  Quantum state estimation by maximum likelihood is biased
  toward more mixed states, and the amount of bias increases with
  increasing purity of the state from which the measurements are drawn
  \cite{glancy2009}.  Based on numerical experiments, the bias in our
  estimates of the fidelity should be well below the uncertainty level
  set by systematic effects.  However, the bias in our estimates of
  the minimum variance of the inferred state could be larger.  To
  attempt to quantify this effect, we simulated measuring and
  performing maximum likelihood tomography on a Gaussian state.  This
  Gaussian state is chosen to have minimum and maximum variances equal
  to those calculated by the linear method described above for the
  ``best-guess'' case.  By computer we simulate 10,000 quadrature
  measurements (the same number we used for ML analysis of the true
  experiment) from this Gaussian state and perform maximum likelihood
  tomography on those measurements.  The inferred state has minimum
  variance $40~\%$. Therefore it is possible that the experimental
  state has smaller minimum variance than the most likely state
  inferred from only 10,000 measurements. Because we have some
  independent evidence for non-Gaussian effects in the experiment, we
  cannot quantify this size of this bias using this Gaussian
  simulation. Other numerical simulations have confirmed that this
  bias decreases as the number of measurements analyzed increases and
  that this bias is not caused by truncation of the Hilbert space at
  30 photons.

The apparent discrepancy between the $6~\%$ for the variance of
$|\psi\rangle$ and the $48~\%$ for the variance of $\rho$ also deserves
some comments.  It is important to note that one would not expect
the minimum variance of a mixed state to equal the minimum variance
of its highest fidelity pure state.  The fidelity between a mixed
Gaussian state (centered at the origin of phase space) whose minimum
and maximum variances are $v_x$ and $v_p$ and a pure squeezed vacuum
state with minimum variance $v_s$ is given by
\[
F_{\mathrm{Gauss}}=\frac{2}{\sqrt{\frac{(1+4v_sv_p)(v_s+v_x)}{v_s}}}.
\]
The highest fidelity pure state has minimum variance
$v_s = \frac{1}{2}\sqrt{\frac{v_x}{v_p}}$, and the fidelity between
these two states is
\[
F_{\mathrm{Gauss,max}}=\frac{2}{1+2\sqrt{v_xv_p}}
\]
Consider the state $\sigma$ to be a Gaussian state with minimum
variance of $48~\%$ and maximum variance $2017~\%$.  ($\sigma$ has
variances equal to those of our state $\rho$, but unlike $\rho$,
$\sigma$ is guaranteed to be Gaussian.)  Then let $|\psi\rangle$ be
the pure squeezed vacuum state that has maximum fidelity with
$\sigma$.  $F_{\mathrm{Gauss,max}} =
\langle\psi|\sigma|\psi\rangle=0.49$, and the minimum variance of
$|\psi\rangle$ is $7.7~\%$.  The difference between the minimum variances of
$\rho$ and $\left|\psi\right\rangle$ is to be expected.  However, the
maximum fidelity of $\rho$ is significantly larger than we would
expect if it was perfectly Gaussian.  This non-Gaussianity could be
caused by bias in the maximum likelihood inference and / or genuine
non-Gaussian effects in the experiment.

Tomographic reconstruction of a quantum state requires that the
experimental device always creates the same (potentially mixed) quantum state,
that the measurements are well described by inefficient quadrature
measurements, and that the calibration of those measurements is
consistent.  In this experiment we have observed some evidence that at
least one of these assumptions is violated.
% When we perform
% tomography under the ``best-guess'' calibration, but not correcting
% for detector inefficiency, we find that the minimum variance of the
% maximum likelihood state is $90.0\pm 0.6~\%$ of the vacuum variance
% (where the uncertainty is only due to statistical effects).  However,
% the minimum variance of the scaled quadrature measurements is only
% $67.9 \pm 0.7~\%$ of the vacuum variance.
The likelihood of the maximum likelihood state is significantly lower than
one should expect from simulated measurements on that state.  That is,
if the tomographic assumptions above were true, we expect to find a
significantly higher value for the maximum likelihood.  We believe
this effect could be caused by an interaction between the state
preparation and measurement stages of the experiment, such as a phase
dependent efficiency of the measurement JPA, and/or non linear
processes in the measurement.

%\bibliography{journal_titles_abbreviated,mybib3}
\end{document}